\documentclass{article}

\usepackage[preprint]{neurips_2025}

\usepackage[utf8]{inputenc} 
\usepackage[T1]{fontenc}    
\usepackage{hyperref}       
\usepackage{url}            
\usepackage{booktabs}       
\usepackage{amsfonts}       
\usepackage{nicefrac}       
\usepackage{microtype}      
\usepackage{xcolor}         
\usepackage{amsmath}
\usepackage{acronym} 
\usepackage{mathtools}
\usepackage{bbm}
\usepackage{subfigure}

\acrodef{COMPAS}{Compact Object Mergers: Population Astrophysics and Statistics}
\acrodef{BH}{black hole}
\acrodef{NS}{neutron star}

\title{RESOLVE: Rare Event Surrogate Likelihood for Gravitational Wave Paleontology Parameter Estimation}

\author{%
  Ann-Kathrin Schuetz$^{1,\dagger}$  \\
  \texttt{aschuetz@lbl.gov}  \\
  \And
  Alexander Migala$^{3,\dagger}$  \\
  \texttt{amigala@ucsd.edu}\\
  \And
  Adam Boesky$^{6}$ \\
  \texttt{aboesky@college.harvard.edu}\\
  \And
  Alan W. P. Poon$^{1,*}$ \\
  \texttt{awpoon@lbl.gov} \\
  \And
  Floor S. Broekgaarden$^{4,*}$ \\
  \texttt{fbroekgaarden@ucsd.edu} 
  \And
  Aobo Li$^{2,3*}$ \\
  \texttt{aol002@ucsd.edu} \\
}

\begin{document}

\maketitle
\begin{center} 
\vspace{-2em}
$^{1}$Nuclear Science Division, Lawrence Berkeley National Laboratory, Berkeley, CA 94720, USA\\
$^{2}$Hal\i c{\i}o\u{g}lu Data Science Institute, UC San Diego, La Jolla, CA  92093, USA\\
$^{3}$Department of Physics, UC San Diego, La Jolla, CA  92093, USA\\
$^{4}$Department of Astronomy and Astrophysics, University of California, San Diego, La Jolla, CA 92093, USA\\
$^{5}$Department of Physics, UC San Diego, La Jolla, CA  92093, USA\\
$^{6}$Center for Astrophysics, Harvard \& Smithsonian, Cambridge, MA 02138, USA\\
$\dagger$ Equal Contribution\\
$^*$ Corresponding Authors
\end{center}

\newcommand{\todo}[1]{\color{red}[#1]\color{black}}
\newcommand{\citecompas}[0]{\cite{compas_team_compas} \cite{stevenson_formation_2017} \cite{vigna-gomez_formation_2018}}
\newcommand{\newparagraph}[0]{\\ \\}
\newcommand{\pgiven}[2]{P(\text{#1}|\text{#2})}

\maketitle

\begin{abstract}
The first detection of gravitational waves, recognized by the 2017 Nobel Prize in Physics, has opened up a new research field: gravitational-wave paleontology. When massive stars evolve into black holes and collide, they create gravitational waves that propagate through space and time. These gravitational-waves, now detectable on Earth, act as fossils tracing the histories of the massive stars that created them. Estimating physics parameters of these massive stars from detected gravitational-waves is a parameter estimation task, with the primary difficulty being the extreme rarity of collisions in simulated binary black holes. This rarity forces researchers to choose between prohibitively expensive simulations or accepting substantial statistical variance. In this work, we present RESOLVE, a rare event surrogate model that leverages polynomial chaos expansion (PCE) and Bayesian MCMC to emulate this rare formation efficiency. Our experimental results demonstrate that RESOLVE is the only surrogate model that achieves proper statistical coverage, while effectively learning the underlying distribution of each physics parameter. We construct a likelihood function incorporating both the emulated formation efficiency and LIGO's gravitational wave observations, which we then minimize to produce community-standard credible intervals for each physics parameter. These results enable astronomers to gain deeper insights into how the universe transformed from simple gases into the complex chemical environment that eventually made life possible.

\end{abstract}

\section{Introduction}
On September 14, 2015, Laser Interferometer Gravitational-Wave Observatory (LIGO) announced the detection of GW150914, marking humanity's first ever direct observation of gravitational waves~\citep{ligo}. This groundbreaking result was swiftly endorsed by the 2017 Nobel Prize in Physics. Gravitational waves were first theorized by Albert Einstein in 1916~\citep{Einstein1916,Einstein1918} as part of his general relativity framework, which predicted that massive objects distort the space and time around them and when they accelerate, they create ``ripples in spacetime'':  gravitational waves that propagate through the universe at the speed of light.

The astronomical events that generate detectable gravitational waves involve the so-called \textbf{binary systems}: when two massive stars evolve into two black holes (or neutron stars), they will orbit around each other at accelerated speed and eventually collide. These collisions release tremendous energy, producing gravitational-waves that can be detected with observatories such as LIGO \cite{ligo}, VIRGO \cite{virgo}, and KAGRA \cite{kagra}~\footnote{See Appendix \ref{app:gravity_radiation_rotation} for more details}. 
Because it can take millions to billions of years for binary black holes to collide and be detected through gravitational waves, these signals can act as ``astronomical fossils'', tracing the histories of millions of their progenitor stars across cosmic time. This new frontier is called Gravitational Wave Paleontology. 

Learning the properties of massive stars in binary system from detected gravitational waves can be described as a parameter estimation task. Suppose the massive stars exhibit certain properties that can be characterized by physics parameters $\theta$: the first step involves a theoretical forward model that accepts $\theta$ and outputs whether a pair of black holes will be formed to create a gravitational wave. This output can then be compared to the actual gravitational wave observations $y_{obs}$ to draw a confidence limit or credible intervals over $\theta$. Astronomers can then use these intervals to test existing theories of the evolution of stars and the underlying physical processes such as stellar outflows, enrichment, nuclear fusion and supernovae, leading to an improved understanding of how the universe transformed from simple gases into the complex chemical environment that eventually made life possible.

The key challenge lies in the theoretical forward model, which, in the astronomy community, is encoded within a well-established software framework Compact Object Mergers: Population Astrophysics and Statistics~(COMPAS) \citecompas. Currently, COMPAS simulations are limited by uncertainty and high computation cost, making it impossible to explore the full parameter space of  $\theta$. Suppose $N$ binary systems are simulated in COMPAS under physics parameter $\boldsymbol{\theta}$, but only $m$ of them form black hole pairs that eventually collide and produce observable gravitational wave signals.  The \textbf{formation efficiency} is thereby calculated as $\epsilon = m/N$. Given that $\epsilon$ is intrinsically small, millions of binary systems must be simulated to collect only a few collisions, making each evaluation of $\epsilon$ computationally prohibitive. Consequently, learning a mapping between $\boldsymbol{\theta}$ and $\epsilon$ becomes intractable under traditional approaches. 

In this work, we developed the RESOLVE model, which includes a novel rare event surrogate model that, after training, maps $\boldsymbol{\theta}$ to $\hat{\epsilon}$ without actually running the time-consuming COMPAS simulation. We can further convert $\hat{\epsilon}$ into an expected gravitational wave rate $\hat{y}$ to be compared to the observed rate $y_{obs}$. The last step involves constructing a likelihood function $L(y_{obs}|\hat{y})$ and minimizing it with Bayesian inference algorithms to obtain credible interval over $\boldsymbol{\theta}$. 


\section{Related Works}~\label{sec:related_works}
It is commonplace across disciplines to use machine learning techniques as surrogates for computationally-expensive simulations \citep{donnelly_physics-informed_2024, wurth_physics-informed_2023, oldenburg_geometry_2022, shibata_novel_2022, sasanapuri_surrogate_2025, fu2024generative}. Most closely related to this work is the RESuM model by \cite{schuetz2025resum}, which addresses similar rare event design problems in physics detector design optimization using a Multi-Fidelity Gaussian Process (MFGP) surrogate model. The key insight is the adoption of Conditional Neural Processes (CNP, \cite{garnelo2018}) to smooth out the discreteness of the rare design metric, providing additional prior information to the MFGP surrogate model. In this work, CNP was also adopted for similar purpose, but we designed a novel surrogate model based on Polynomial Chaos Expansion (PCE)~\cite{xiu2002} for binary black hole simulation.

Another related field is rare event simulation and modeling in reliability engineering. The rare event problem here focuses on emulating the extremely low failure probabilities $P_f$. Since direct Monte Carlo simulation becomes intractable as $P_f$ approaches zero, specialized techniques including adaptive sampling~\citep{bucher1988adaptive}, surrogate-based methods~\citep{li2010evaluation,li2011efficient}, sequential importance sampling~\citep{papaioannou2016sequential}, and multi-fidelity approaches~\citep{peherstorfer2016multifidelity,peherstorfer2018multifidelity}, multilevel sampling~\citep{wagner2020multilevel} and ensemble Kalman filters~\citep{wagner2022ensemble} are developed. While Adaptive Importance Sampling~(AIS) method has the potential to solve the rare event problem,  the benchmarking study in \cite{schuetz2025resum} showed that it is difficult to implement yet still underperform the CNP approach. It is worth noting that an AIS method called STROOPWAFFLE~\cite{broekgaarden_stroopwafel_2019} has been implemented in COMPAS for simulations and can improve the initial condition sampling but has not been used as a surrogate model.

Lastly, in particle physics and astrophysics, parameter estimation relies extensively on both frequentist and Bayesian statistical inference. The chi-square fitting method, a cornerstone of frequentist inference, has been employed for decades to extract physical parameters from experimental data, particularly in collider experiments \cite{cowan2011statistical} and neutrino oscillation studies \cite{feldman1998unified}. Complementarily, Bayesian inference utilizing Markov Chain Monte Carlo (MCMC) methods has gained significant attention due to its ability to incorporate prior knowledge \cite{trotta2008bayes, feroz2009multinest, straub2024inverse}. These methods have proven especially valuable in cosmological parameter estimation \cite{planck2020parameters}, dark matter searches \cite{XENON:2023cxc, LZ:2024zvo}, and the search for rare decay processes \cite{KamLAND-Zen:2022tow, GERDA:2020xhi, Majorana:2022udl}. In RESOLVE, we inherited the conventional approaches from physics and astronomy to construct the likelihood function and draw credible interval. We also provided a more detailed description about Bayesian inference in Physics in Appendix~\ref{sec:bmcmc} and~\ref{app:Bmcmc_particle_physics}



\section{The RESOLVE Model}\label{section:model}
RESOLVE is a rare event surrogate likelihood constructed to solve the rare event problem Gravitational Wave Paleontology parameter estimation. A theoretical formalization of the rare event problem can be found in Appendix~\ref{sec:theory}. Consider a scenario where we run $J$ simulation trials indexed by \(j\); each simulation trial contains \(N\) binary systems indexed by $i$. For each system $i$ in trial $j$, The outcome $X_{ji}$ is either 1, meaning that the $i^{th}$ pair of stars formed black hole pairs that collided and emitted a detectable gravitational wave, or 0, meaning that the binary system did not form colliding black holes. 
If $m$ black hole formed in $N$ simulated binary systems, the formation efficiency $\epsilon$ for trial $j$ can be defined as:
\begin{equation}
    \epsilon_j = \frac{m}{N} = \frac{\sum_{i=1}^{N} X_{ji}}{N}
    \label{eqn:yhat}
\end{equation} 
All events in the same trial have the same value of the parameter of interest \(\boldsymbol{\theta}_j\) but a different nuisance parameter $\boldsymbol{\phi}_{ji}$. Each trial can be either Low-Fidelity (LF) or High-Fidelity (HF), depending on the number of events simulated (\(N_{HF} \gg N_{LF}\)). The LF simulation suffers from high statistical variance since $N_{LF}$ is small, but the lower computational cost enables broader exploration of the \(\boldsymbol{\theta}\) parameter space. Meanwhile, the more expensive HF simulations allow us to obtain a $\epsilon$ that is a better estimate of the true formation efficiency. In the following context, we define $\epsilon_{Raw}$ as the formation efficiency calculated directly from $m/N$.


This section provides a comprehensive overview of the design and training of RESOLVE model. The first step involves training a Conditional Neural Process (CNP) model to produce CNP scores $\beta_{ji}$ for all simulated events. Section~\ref{subsec:MFBPCE} describes our newly developed surrogate model based on the Multi-Fidelity Bayesian Polynomial Chaos Expansion (MF-BPCE). This model is trained using both averaged CNP scores and raw formation efficiency measurements. After training, the surrogate model maps $\boldsymbol{\theta}$ to estimated formation efficiency with quantified uncertainty: $\boldsymbol{\theta}\xrightarrow[]{}\hat{\epsilon}\pm\hat{\sigma}$. Finally, Section~\ref{subsec:likelihood} describes the construction of a likelihood function that connects the emulated efficiency values $\hat{\epsilon}$ with observed gravitational wave signals $y_{obs}$.

\subsection{Multi-Fidelity Bayesian Polynomial Chaos Expansion}\label{subsec:MFBPCE}

\begin{figure}[hbt!]
    \centering
    \includegraphics[width=0.99\linewidth]{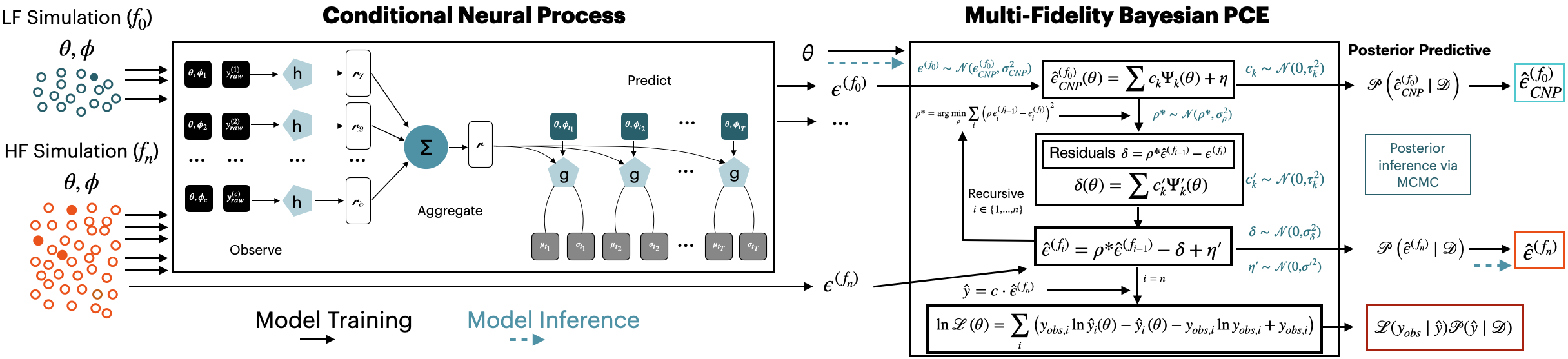}
    \caption{Overview of the RESOLVE framework. The left side illustrates the CNP used for modeling both LF and HF simulation data. The CNP aggregates nuisance parameters \(\boldsymbol{\phi}_i\) and parameters of interest \(\boldsymbol{\theta}\) from LF and HF simulations to produce \(\epsilon_\text{CNP}^{(f_0)}\), which, together with HF formation efficiency \(\epsilon^{(f_n)}\), serve as inputs to the surrogate model. The right side shows the multi-MFGP that combines predictions $\hat{\epsilon}_\text{CNP}^{(f_0)}$ from LF and HF to estimate the HF formation efficiency \(\hat{\epsilon}^{(f_n)}\).}
    \label{fig:overview_mfpce}
\end{figure}

The model integrates sparse Polynomial Chaos Expansion (PCE) with hierarchical Bayesian inference to efficiently propagate uncertainty across both HF and LF simulations trials. The goal is to emulate the raw formation efficiency $\epsilon^{HF}_{Raw}=m/N$ evaluated on HF simulations, using both LF and optionally medium-fidelity (MF) data, thereby reducing computational cost while maintaining accuracy and uncertainty quantification capabilities. Figure \ref{fig:overview_mfpce} shows a schematic overview of the model approach.

Let \( \boldsymbol{\theta} \in \mathbb{R}^l \) be the input parameter vector, and let \( \epsilon^{(f_i)}(\boldsymbol{\theta}) \in \mathbb{R} \) denote the model output at fidelity level \( f_i \in \mathcal{F} = \{f_0, f_1, \dots, f_n\} \), where \( f_0 \) is the lowest and \( f_n \) the highest fidelity. The goal is to construct a surrogate for \( \epsilon^{(f_n)}(\boldsymbol{\theta}) \) that uses all available fidelity data and quantifies uncertainty in its predictions. \( \epsilon^{(f_0)}(\boldsymbol{\theta}) \) can be expanded into a finite series of multivariate orthogonal polynomials:
\[
\hat{\epsilon}^{(f_0)}(\boldsymbol{\theta}) = \sum_{k=1}^{d^{(f_0)}} c^{(f_0)}_k \Psi_k(\tilde{\boldsymbol{\theta}}),
\]
where \( \tilde{\boldsymbol{\theta}} \in [-1,1]^d \) is the normalized input,and \( \Psi_j(\cdot) \) are multivariate Legendre basis functions formed as tensor products of 1D polynomials:
\[
\Psi_\alpha(\tilde{x}) = \prod_{i=1}^d L_{\alpha_i}(\tilde{x}_i),
\]
with multi-index \( \alpha \in \mathbb{N}_0^d \) satisfying \( \|\alpha\|_1 \leq d^{(f_0)} \), and \( d^{(f_0)} \) being the maximum total degree.

For the multi-fidelity surrogate construction, we adopt an autoregressive model inspired by the work~\cite{kennedy2000}. In this framework, each fidelity level \( f_i \in \{f_0, \dots, f_n\} \) is assumed to have a linear (scalar) relationship with the prediction from the preceding fidelity level \( f_{i-1} \), plus an additive discrepancy term modeled via a PCE as well. Formally, the model is defined recursively as
\begin{align}
\hat{\epsilon}^{(f_i)}(\boldsymbol{\theta}) &= \rho^{(f_i)} \hat{\epsilon}^{(f_{i-1})}(\boldsymbol{\theta}) + \sum_{k=1}^{d^{(f_i)}} c_k^{(f_i)} \Psi_k(\tilde{\boldsymbol{\theta}}), \quad \text{for } f_i = f_0, \dots, f_n.
\end{align}
where higher-fidelity predictions \( \hat{\epsilon}^{(f_i)}(\boldsymbol{\theta}) \) are expressed as corrections to the scaled lower-fidelity predictions with \( \rho^{(f_i)} \in \mathbb{R} \) being a scalar scaling parameter capturing the linear relationship between fidelities \( f_i \) and \( f_{i-1} \). The discrepancy term \( \sum_k c_k^{(f_i)} \Psi_k(\tilde{\boldsymbol{\theta}}) \) captures residual structure not accounted for by the scaled prediction \( \rho^{(f_i)} \hat{\epsilon}^{(f_{i-1})}(\boldsymbol{\theta}) \), allowing the model to learn fidelity-specific corrections. 

This autoregressive multi-fidelity structure naturally enables coherent uncertainty propagation across fidelity levels by explicitly modeling the relationships between low- and high-fidelity outputs within a probabilistic framework. To formalize this, we adopt a Bayesian linear regression formulation that allows us to quantify and propagate epistemic uncertainty in the model coefficients and discrepancy terms. Specifically, we treat the model output \( \epsilon^{(f_i)} \) at fidelity level \( f_i \) as a noisy observation of a PCE
\[
\epsilon^{(f_i)} \sim \mathcal{P}\left(\hat{\epsilon}^{(f_i)}(\boldsymbol{\theta}_j), \boldsymbol{\eta}^{(f_i)}\right),
\]

where \( \hat{\epsilon}^{(f_i)}(\boldsymbol{\theta}_j) \) denotes the deterministic PCE approximation at input \(\boldsymbol{\theta}_j \), and \( \boldsymbol{\eta}^{(f_i)} \) are parameters governing the noise or discrepancy distribution at fidelity level \( f_i \). The likelihood function \( \mathcal{P} \) can be chosen flexibly, depending on the characteristics of the data. 

To further reduce the statistical noise in the low-fidelity outputs, \cite{schuetz2025resum} leveraged a CNP algorithm to transform binary outcomes of $X_{ji}$ (collision or no collision) into continuous, floating-point scores $\beta_{ji}$ between 0 and 1. In this framework, the CNP learns a Gaussian approximation to the posterior of the latent function:
\[
q(t(\boldsymbol{\theta},\boldsymbol{\phi})) \approx \mathcal{N}(\mu_{\text{NN}}(\boldsymbol{\theta},\boldsymbol{\phi}), \sigma^2_{\text{NN}}(\boldsymbol{\theta},\boldsymbol{\phi})),
\]
trained by minimizing the expected negative log-likelihood under the Bernoulli model:
\[
\mathcal{L} = - \sum_{k=1}^{K} \sum_{i=1}^{N_k} \log \int \text{Bernoulli}(X_{ki} \mid t(\boldsymbol{\theta}_k, \boldsymbol{\phi}_{ki})) \, q(t(\boldsymbol{\theta},\boldsymbol{\phi})) \, dt.
\]
The predicted mean \( \mu_{\text{NN}} \) serves as a smoothed estimate \( \beta_{ki} \approx t(\boldsymbol{\theta}_k, \boldsymbol{\phi}_{ki}) \). These scores are averaged across instances to produce the low-fidelity design metric:
\begin{equation}
    \epsilon^{(f_0)}_{\text{CNP}} = \frac{1}{N_k} \sum_{i=1}^{N_k} \beta_{ki}.
    \label{eqn:cnp_score}
\end{equation}
This smooth surrogate \( \epsilon^{(f_0)}_{\text{CNP}} \) replaces the direct empirical estimate \( m/N \), providing a denoised input for downstream surrogate modeling. Benchmarking results in \cite{schuetz2025resum} demonstrated that including the CNP scores enhances the performance of Multi-Fidelity Gaussian Process (MFGP) surrogate models. In this work, we adopted the same CNP structure as \cite{schuetz2025resum} for the gravitational wave analysis. It acts as a learned prior for the lowest-fidelity level, and its predictions are passed into the autoregressive multi-fidelity model:

\[
\epsilon^{(f_0)} \sim \mathcal{N}\left(\epsilon^{(f_0)}_{\text{CNP}}, \sigma^2_{\text{CNP}}\right).
\]

The coefficients of the PCE, denoted \( \{c_k^{(f_i)}\} \), define the contribution of each basis function at fidelity level \( f_i \), while a scaling parameter \( \rho^{(f_i)} \) modulates the influence of the lower-fidelity prediction. These parameters are treated as latent parameters with prior distributions:
\[
\{c_k^{(f_i)}\} \sim \mathcal{P}_{\text{c}}, \quad \boldsymbol{\eta}^{(f_i)} \sim \mathcal{P}_{\text{n}}, \quad \rho^{(f_i)} \sim \mathcal{P}_{\text{s}}.
\]
The prior on \( \{c_k^{(f_i)} \}\) regularizes the expansion to avoid overfitting. The specific choice of priors \( \mathcal{P}_{\text{c}}, \mathcal{P}_{\text{n}}, \mathcal{P}_{\text{s}} \) can be adapted to reflect prior knowledge, modeling needs, or regularization preferences.

Given data \( \mathcal{D}^{(f_i)} = \{(\boldsymbol{\theta}_j, \epsilon_j^{(f_i)})\}_{j=1}^N \), we infer the posterior distribution over the latent parameters via Bayes' rule:
\[
\mathcal{P}(\{c^{(f_i)}\}, \rho^{(f_i)}, \boldsymbol{\eta}^{(f_i)} \mid \mathcal{D}^{(f_i)}) \propto \mathcal{L}(\mathcal{D}^{(f_i)} \mid \{c^{(f_i)}\}, \rho^{(f_i)}, \boldsymbol{\eta}^{(f_i)}) \cdot \mathcal{P}_{c}(\{c^{(f_i)}\}) \cdot \mathcal{P}_{s}(\rho^{(f_i)}) \cdot \mathcal{P}_{n}(\boldsymbol{\eta}^{(f_i)}),
\]
where \( \mathcal{L} \) denotes the likelihood function induced by the assumed output distribution \( \mathcal{P} \).

For a new input \(\boldsymbol{\theta}^* \), the model yields a posterior predictive distribution:
\[
\mathcal{P}(\hat{\epsilon}^* \mid \boldsymbol{\theta}^*, \mathcal{D}) = \int \mathcal{P}(\hat{\epsilon}^* \mid \boldsymbol{\theta}^*, \boldsymbol{\vartheta}) \, \mathcal{P}(\boldsymbol{\vartheta} \mid \mathcal{D}) \, d\boldsymbol{\vartheta},
\]
where \( \boldsymbol{\vartheta} \) denotes the full set of model parameters. The predictive distribution captures both epistemic uncertainty due to limited data and aleatoric uncertainty inherent in the model formulation.

The MF-BPCE surrogate model algorithm maps an input $\boldsymbol{\theta}$ to both $\epsilon^{LF}_{CNP}$ as low-fidelity output and $\epsilon^{HF}_{Raw}$ as high-fidelity outputs via it's posterior predictive:
\begin{equation}
f:\boldsymbol{\theta}\xrightarrow {} {\mathcal{P}(\hat{\epsilon} \mid \mathcal{D})}
\end{equation}

\subsection{Likelihood Construction}\label{subsec:likelihood}
It is worth noting that the emulated formation efficiency $\hat{\epsilon}$ is still different from the expected gravitational wave rate $\hat{y}$. While a more accurate conversion exists in COMPAS, implementing it is not straightforward, so we decided to leave it for future work. In this work, we adopted a simplified conversion using a constant multiplication factor: $\hat{y} = 314.266854 \cdot \hat{\epsilon}$, which is consistent with community standards. Details regarding this conversion factor can be found in Appendix~\ref{app:epsilon_to_y}.


We then constructed a likelihood function with both $\hat{y}$ and $y_{obs}$. Because the observed number of events in a fixed volume and time interval is a discrete count, and the events are assumed to occur independently and rarely, the Poisson distribution is the natural and widely adopted model in the astroparticle physics community for such data~\citep{cousins2018lectures}. Thus, we define the log-likelihood as:

\begin{equation}
    \ln \mathcal{L}(\boldsymbol{\theta}) =  \sum_i \left(y_{obs,i} \ln\hat{y_i}(\boldsymbol{\theta}) - \hat{y_i}(\boldsymbol{\theta}) - \ln y_{obs,i}\mathbb{!} \right)
\end{equation}

To infer the posterior over the model parameters $\boldsymbol{\theta}$, we integrate the Poisson likelihood over the surrogate model output $\hat{y}$, and over the latent model parameters $\boldsymbol{\vartheta}$:

\[
\mathcal{P}(\boldsymbol{\theta}\mid y_{obs},\mathcal{D}) = \int \int \mathcal{L}(y_{obs}\mid \hat{y}) \mathcal{P}(\hat{y} \mid \boldsymbol{\theta}, \boldsymbol{\vartheta}) \, \mathcal{P}(\boldsymbol{\vartheta} \mid \mathcal{D}) \, d\hat{y}\, d\boldsymbol{\vartheta},
\]

This formulation reflects both epistemic uncertainty (via $\boldsymbol{\vartheta}$) and aleatoric uncertainty (via $\hat{y}$) in the final inference over $\boldsymbol{\theta}$. 

\section{Experiment and Validation}\label{section:experiment}
Given that the likelihood function relies primarily on $\hat{\epsilon}$, it is crucial to validate that the surrogate model produces accurate estimates of $\hat{\epsilon}\pm\hat{\sigma}$. This section presents experimental results benchmarking RESOLVE against other surrogate models on two different datasets. The three models include \textbf{MFGP}---the Multi-Fidelity Gaussian Process surrogate model without the use of Conditional Neural Process~(CNP); \textbf{RESuM}---the Rare Event Surrogate Model proposed in~\cite{schuetz2025resum}, including both MFGP and the CNP; and the \textbf{RESOLVE} model proposed in this work includes the CNP and the MF-BPCE algorithm described in Section~\ref{subsec:MFBPCE}. The two datasets include the LEGEND detector design dataset and the binary black hole collision dataset. Details on the concrete implementation and diagnostic are provided in Appendix~\ref{sec:diagnostics}.

The LEGEND detector design dataset was provided by \cite{schuetz2025resum} with a similar HF/LF structure. The overall goal is to surrogate the design metric $\hat{p}=m/N$ from given design parameter $\boldsymbol{\theta}$. More details about this dataset are provided in Appendix~\ref{app:LEGEND_data} and the original paper. Benchmarking was performed on a separate validation dataset with 100 out-of-sample HF simulation trials. Each models were benchmarked with 4 metrics: the Means Square Error (MSE) calculated by averaging $(\hat{p}-p_{Raw})^2$ over the 100 trials, and the 1$\sigma$, 2$\sigma$, and 3$\sigma$ coverage calculated by counting the percentage of trials where $p_{Raw}$ falls within $\hat{p}\pm1\hat{\sigma}$, $\hat{p}\pm2\hat{\sigma}$, and $\hat{p}\pm3\hat{\sigma}$, respectively. As shown in Table~\ref{tab:benchmark}, the MFGP algorithm (Trial 1) is significantly undercovered. The RESuM model (Trial 2) and the RESOLVE model (Trial 3) both achieved proper statistical coverage on the LEGEND detector design dataset, while RESOLVE outperforms RESuM on the MSE metric.

The major scope of this work is the black hole collision dataset we generated from scratch with COMPAS v03.10.05. For each simulated binary system in COMPAS, users can specify 27 input parameters listed in Appendix~\ref{app:param_choices}. In this work, we selected 4 out of 27 to form the physics parameter of interest $\boldsymbol{\theta}$: the $metallicity$ parameter $Z$ that determines the initial fraction of metals that the stars are made of; the common-envelope  efficiency parameter $\alpha_{CE}$ that impacts the total shrinking of the binary's orbit during an unstable mass transfer episode; the $\sigma_{BH}$ parameter which represents the MSE of the Maxwellian distribution from which black hole natal kicks are sampled during a supernova event, and the $\sigma_{NS}$ parameter which is the same as $\sigma_{\rm{BH}}$ but for neutron stars. The other 23 parameters are treated as the nuisance parameter $\boldsymbol{\phi}$.

We then generated HF and LF simulations trials under different $\boldsymbol{\theta}$. Each HF trial cost 4 CPU hours to generate $10^{6}$ binary systems, while each LF trial cost 15 CPU seconds to generate $10^{3}$ of them. Within in each trial, the 4 $\boldsymbol{\theta}$ parameters are kept constant throughout all simulated binary systems, while the other 23 $\boldsymbol{\phi}$ parameters vary randomly among different systems. The formation efficiency is thus obtained as $\epsilon_{RAW} = m/N$, as described earlier. The training dataset contains 1000 LF trials and 15 HF trials, while the validation dataset contains 150 out-of-sample HF simulation trials.

\begin{figure}[hb!]
    \centering
    \includegraphics[width=0.9\linewidth]{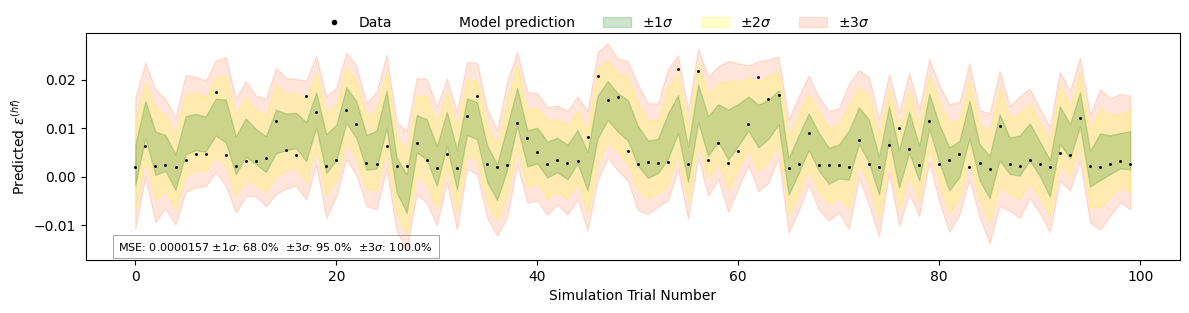}
    \caption{Coverage plot of the RESOLVE model predictions on the binary black hole dataset (Trial 14 in Table~\ref{tab:benchmark}).}
    \label{fig:validation_bbh}
\end{figure}

\begin{table}[ht!]
\caption{Benchmarking Result on both the LEGEND Detector Design Dataset and the Binary Black Hole Dataset. For RESOLVE model, the (d=2/3/4) in parenthesis means the choice of $2^{nd}/3^{rd}/4^{th}$ order polynomial of the Polynomial Chaos Expansion algorithm, respectively.}
\label{tab:benchmark}
\begin{center}
\footnotesize
\begin{tabular}{cccccccc} \hline
\hline
Trial & Model  & Dataset & (\#LF,\#HF) & MSE & 1$\hat{\sigma}$ [\%] & 2$\hat{\sigma}$ [\%] & 3$\hat{\sigma}$ [\%] \\
\hline
1&MFGP \tiny{\cite{schuetz2025resum}}&Detector Design& (310,10)&0.015&17\% &33\% &48\% \\
2&RESuM \tiny{\cite{schuetz2025resum}}&Detector Design& (310,10)&0.002&69\% &95\% &100\% \\
3&RESOLVE&Detector Design &(310,10)&0.001&74\%&94\%&99\%\\
\hline
4&MFGP&Black Hole &(1000,5)  & 8.4$\times10^{-6}$ & 70\% & 89\% & 96\% \\
5&MFGP&Black Hole &(1000,10) & 2.1$\times10^{-6}$ & 47\% & 62\% & 74\% \\
6&MFGP&Black Hole &(1000,15) & 2.3$\times10^{-6}$ & 59\% & 64\%&  72\% \\
\hline
7&RESuM&Black Hole &(1000,5)  &  26.0$\times10^{-6}$ & 70\% & 89\% & 96\% \\
8&RESuM&Black Hole &(1000,10) &  2.2$\times10^{-6}$ & 53\% & 72\% & 81\% \\
9&RESuM&Black Hole &(1000,15) &  2.3$\times10^{-6}$ & 66\% & 74\% & 80\% \\
\hline
10&RESOLVE (d=2)&Black Hole &(1000,15)&12.7$\times10^{-6}$&76\%&97\%&100\%\\
11&RESOLVE (d=3)&Black Hole &(1000,15)&13.1$\times10^{-6}$&74\%&97\%&100\%\\
\hline

12&RESOLVE (d=4)&Black Hole &(1000,5) & 16.1$\times10^{-6}$ & 68\% & 95\% & 100\% \\
13&RESOLVE (d=4)&Black Hole &(1000,10) & 12.6 $\times10^{-6}$& 83\% & 100\% & 100\% \\
14&RESOLVE (d=4)&Black Hole &(1000,15) & 13.0$\times10^{-6}$ & 74\% & 97\% & 100\% \\
\hline

\multicolumn{3}{c}{Proper Statistical Coverage }&&&68.27\%&95.45\%&99.73\%\\
\hline
\hline
\end{tabular}
\end{center}
\end{table}
The benchmarking metrics for the binary black hole datasets are analogous to those of the LEGEND detector design dataset. The results are shown in Table~\ref{tab:benchmark}. Each model was trained with 700 LF trials along with 5/10/15 HF trials. As shown in Trials 3-6, the MFGP model without CNP exhibits significant overcoverage in its predictions; The RESuM Model (MFGP+CNP) in Trials 7-9 reached proper coverage at $1\sigma$ but is significantly undercovered at $2\sigma$ and $3\sigma$. RESOLVE (Trials 10-14) was the only model that achieved proper statistical coverage on binary black hole dataset. We further examined different choices of polynomial order in the Polynomial Chaos Expansion algorithm of RESOLVE, finding that 4$^{th}$ order polynomials yield optimal results. The prediction accuracy and coverage results for each individual trials are illustrated in Figure~\ref{fig:validation_bbh}.

\begin{figure}[h]
    \centering
    \includegraphics[width=0.8\linewidth]{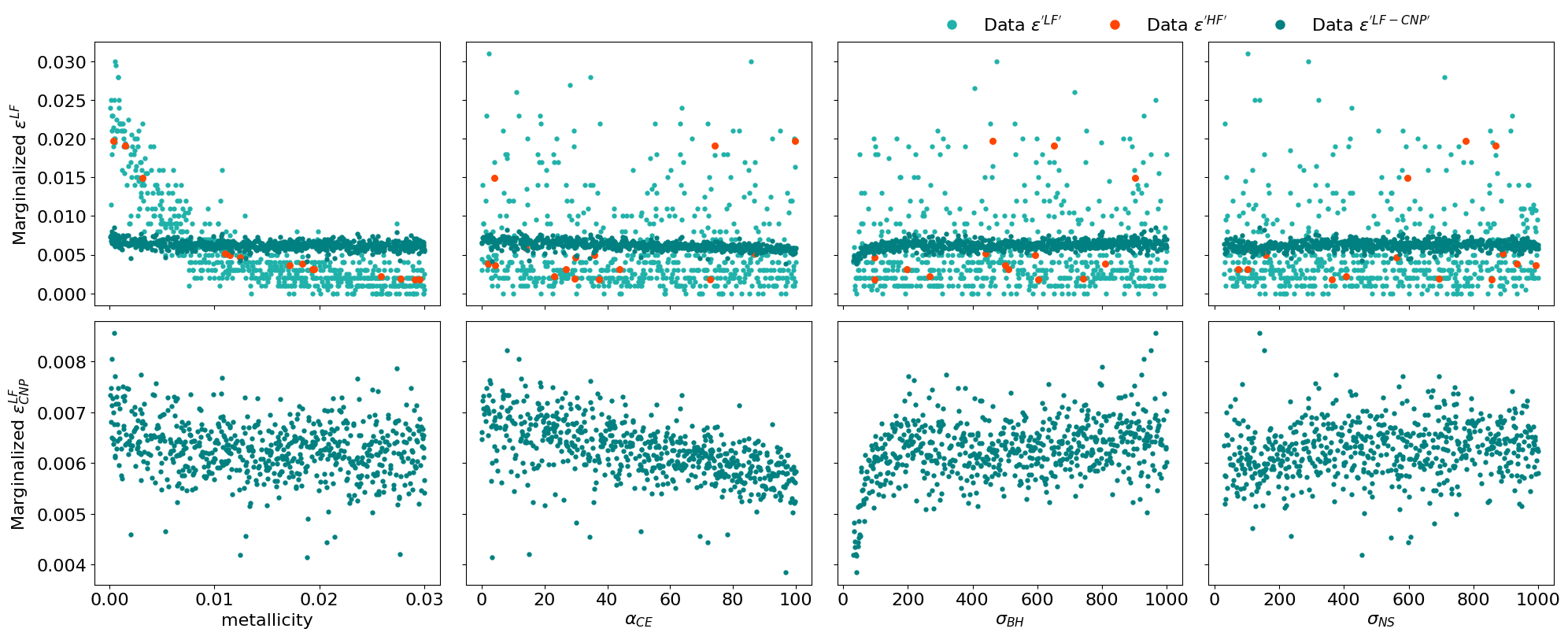}
    \includegraphics[width=0.8\linewidth]{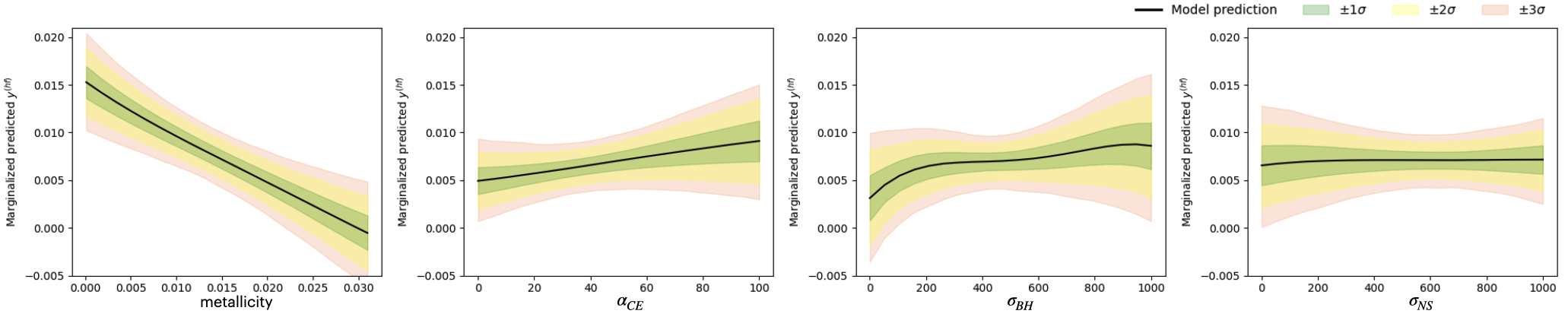}
    \includegraphics[width=0.8\linewidth]{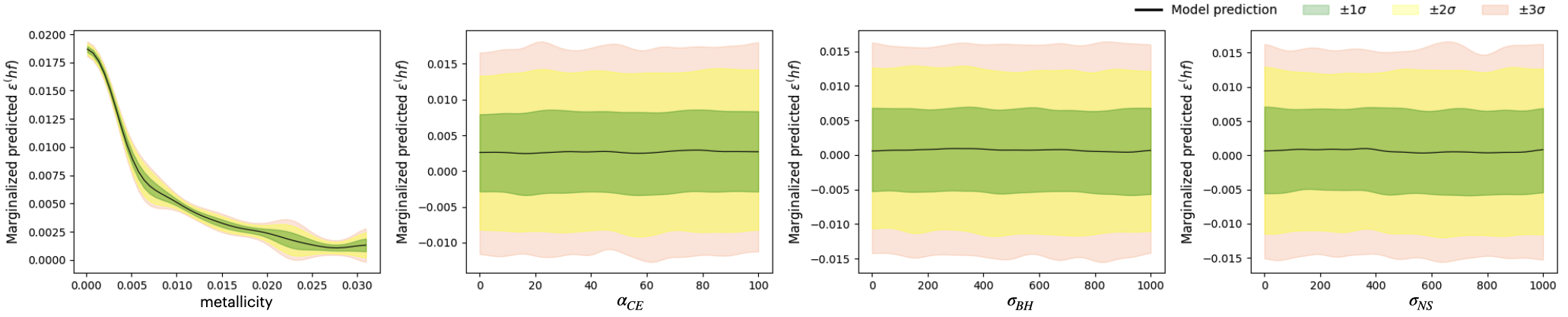}
    \caption{Marginalized distribution of four $\theta$ parameters using different versions of formation efficiency $\epsilon$ as y axis: \textbf{Row 1} (from top) uses $\epsilon^{LF}_{Raw} = m/N$ (light teal) and high-fidelity $\epsilon^{HF}_{Raw} = m/N$ (red) outputs, along with the averaged CNP score $\epsilon^{LF}_{CNP}$(dark teal, see Eqn.~\ref{eqn:cnp_score}); \textbf{Row 2} is a zoomed-in version of $\epsilon^{LF}_{CNP}$ in Row 1; \textbf{Row 3} uses the emulated $\hat{\epsilon}$ from the RESOLVE model; \textbf{Row 4} uses the emulated $\hat{\epsilon}$ from the RESuM model.}
    
    \label{fig:marginalized_distribution}
\end{figure}

According to Table~\ref{tab:benchmark}, the RESOLVE model yields the highest mean squared error (MSE) among all models. Although this might initially seem disadvantageous, a closer examination reveals that RESOLVE is immune to a common limitation from the other surrogate models. Figure~\ref{fig:marginalized_distribution} Row 1 displays the raw formation efficiency $\epsilon_{Raw} = m/N$ calculated on both LF and HF simulations. As expected, the LF simulations exhibit high noise levels that obscure underlying statistical trends. As discussed in Section~\ref{section:model}, the CNP-processed efficiency $\epsilon^{LF}_{CNP}$ calculated using Equation~\ref{eqn:cnp_score} substantially reduces this noise, allowing small-scale trends to emerge from the statistical background.

$\epsilon_{Raw} = m/N$ from HF $\epsilon^{LF}_{CNP}$ from LF were then used to train RESOLVE and RESuM, with the training result displayed in Figure~\ref{fig:marginalized_distribution} Row 3 and 4, respectively. Based on Row 4, it becomes evident that RESuM resorts to rely solely on metallicity $Z$ to predict $\hat{\epsilon}$. Despite experimenting with various kernels, hyperparameter settings, and normalization schemes, the GP regression of the RESuM Model consistently produced nearly flat trend for varied $\alpha_{CE}$, $\sigma_{BH}$, and $\sigma_{NS}$. We attribute this to the combination of the CNP’s smooth output, which dampens local variation, and the sparsity of HF data, which limits the GP’s ability to resolve subtle dependencies. The underlying physical effects—such as those from, $\alpha_{\text{CE}}$, and natal kick dispersion—are often low in amplitude and may fall below the GP’s sensitivity threshold in this context. 

A direct consequence of the flat $\alpha_{CE}$, $\sigma_{BH}$, and $\sigma_{NS}$ trend is the undercoverage of RESuM as shown in Table~\ref{tab:benchmark}. From an astrophysical perspective~\citep[][]{compas_team_compas, Broekgaarden:2022, Boesky:2024}, while metallicity is indeed the dominant factor influencing the binary black hole formation probability $\epsilon$, the remaining three parameters in $\boldsymbol{\theta}$ also have subtle but important effects, especially in rare edge cases. RESuM effectively capture the central trend driven by metallicity to yield good coverage at the $1\sigma$ level. However, it fail to account for the combined influence of $\alpha_{CE}$, $\sigma_{BH}$, and $\sigma_{NS}$, leading to significant undercoverage at the $2\sigma$ and $3\sigma$ levels. This exemplifies a common limitation in surrogate models that prioritize strong predictors while underestimating the cumulative impact of secondary parameters, particularly in boundary regions where these effects become magnified. The consequence is a model that appears well-calibrated for common scenarios but significantly underestimates the probability of rare events—events that are often critical in Gravitational Wave Paleontology.

In contrast, the MF-PCE approach in RESOLVE successfully recovers these parameter dependencies as shown in Figure~\ref{fig:marginalized_distribution} Row 3. This suggests that the relevant functional relationships are globally smooth and can be more effectively captured by the global basis functions of the PCE. Even with sparse data, the PCE is able to detect low-amplitude with consistent trends. We further examined that the subtle trend learned by RESOLVE makes astrophysical sense: For example, we observe a 
a non-monotonic trend with respect to the black hole natal kick dispersion $\sigma_{\text{BH}}$, where the probability initially increases before decreasing. This non-monotonic behavior aligns with physical expectations: moderate kicks can enhance binary disruption, whereas high kicks often eject systems that would not merge in any case \citep[][]{compas_team_compas, Broekgaarden:2022, Boesky:2024}. For $\alpha_{\text{CE}}$, the MF-PCE model predicts a rising trend that differs from the low-fidelity behavior; this is driven by the small number of high-fidelity training points and their localized influence. A detailed analysis of this effect, including its implications and mitigation strategies, is provided in Appendix~\ref{sec:alpha_behavior}. The ability to learn the complex, high-dimensional interplay among parameters is crucial for Gravitational Wave Paleontology and enabled RESOLVE to become the only model that achieved proper coverage across the full range of uncertainty intervals.

\section{Bayesian Inference and Credible Interval}\label{section:physics_result}

%

In Section~\ref{section:experiment}, we demonstrated that RESOLVE produces accurate estimations of formation efficiency with robust statistical coverage while capturing subtle trends across all parameters. Given that the LIGO-Virgo collaboration reported an observed gravitational wave rate of $y_{obs}=17^{+10}_{-6.7}$ yr$^{-1}$ Gpc$^{-3}$ in~\cite{gwtc3:2023}, we ran Bayesian inference to obtain credible intervals for the four physics parameters in $\boldsymbol{\theta}$. The MF-BPCE algorithm in RESOLVE generates a Bayesian posterior during training, which can be directly used as the prior for the Bayesian inference. This approach inherently incorporating uncertainty within each parameter of $\boldsymbol{\theta}$. The Bayesian inference uses MCMC algorithm to sample the $\boldsymbol{\theta}$ space where the likelihood functions is minimized. Additional details on our Bayesian inference methodology can be found in Appendix~\ref{sec:bmcmc}.

\begin{figure}[h]
    \caption{Posterior distributions and correlations of metallicity, $\alpha_{\mathrm{CE}}$, $\sigma_{\mathrm{BH}}$, and $\sigma_{\mathrm{NS}}$, inferred with the MF-PCE algorithm and MCMC. Contours show 68\% and 95\% credible regions.}
    \includegraphics[width=1.0\linewidth]{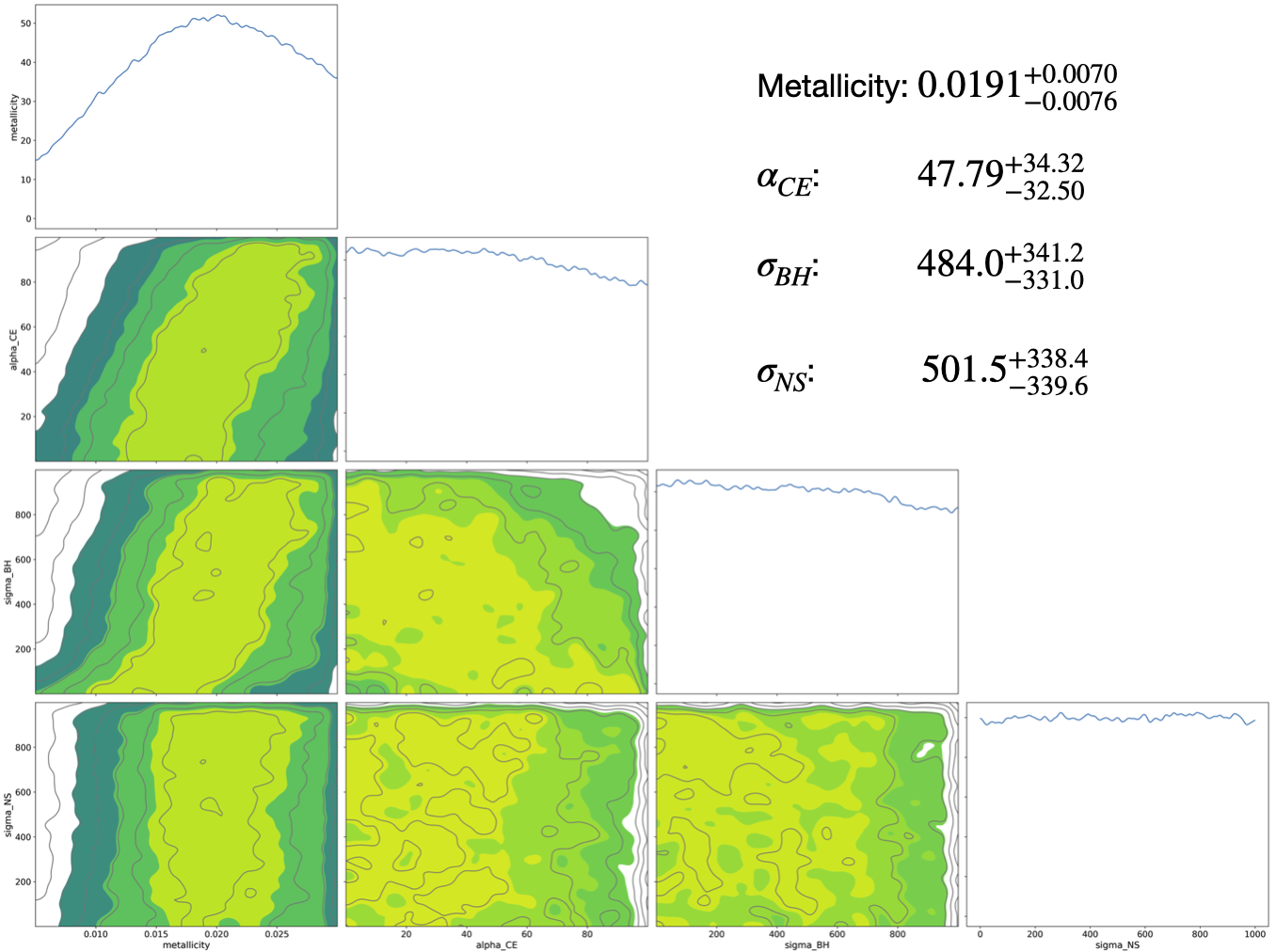}
    \label{fig:resolve_inference}
    \vspace{-3em}
\end{figure}

The results of our Bayesian inference are presented in Figure~\ref{fig:resolve_inference}. Based on the 17 gravitational wave observations, the inference placed a strong constraint on the metallicity parameter with small uncertainties. This precision stems from the strong and clear dependency between metallicity and formation efficiency, leading to a final metallicity value of $0.0191^{+0.007}_{-0.076}$. In contrast, for the remaining three parameters, our model cannot place any strong constraint due to their subtle relationships with formation efficiency, thereby the uncertainty is large. 

\section{Limitations and Applications}\label{section:limitation}

\paragraph{Limitations:} This work has three limitations: (1) the relationship $\hat{y} = 314.266854 \cdot \hat{\epsilon}$ is oversimplified. COMPAS has a more sophisticated Cosmic Integration processer to provide a better mapping from $\hat{\epsilon}$ to $\hat{y}$. (2) The Bayesian inference described in Section~\ref{section:physics_result} failed to draw tight conclusions on three parameters beyond metallicity.(3) The current high-fidelity (HF) simulation which requires $\mathcal{O}(4h)$ computation time is still not the most accurate simulation for binary black holes. A more complex exist but would require $\mathcal{O}(month)$ to generate.
\paragraph{Application:} RESOLVE's capability to learn subtle but important trends and provide credible intervals for user-selected parameters of interest could help to map out, for the first time, the high-dimensional parameter space of these gravitational-wave paleontology simulations. This will disrupt the field by: (i) effectively testing different variations of cosmological model assumptions (ii) studies related phenomena, including supernovae, enrichment, galaxy evolution, instrumentation design, and cosmology; (iii) discover and quantify important trends in  gravitational-wave paleontology simulations such as how the binary black hole formation efficiency depends on metallicity, supernovae physics, mass transfer events, and stellar winds. In the future, we will leverage RESOLVE to explore these research directions to generate more significant scientific results in gravitational-wave paleontology.

\section{Conclusion}\label{section:conclusion}
In this work, we presented RESOLVE, a rare event surrogate likelihood designed for parameter estimation in Gravitational Wave Paleontology. At its core, we created a novel rare event surrogate model based on the Multi-Fidelity Bayesian Polynomial Chaos Expansion~(MF-BPCE) algorithm, which efficiently emulates the binary black hole merger formation  efficiency in gravitational-wave paleontology simulations. We benchmarked RESOLVE against other surrogate models on simulated out-of-sample data, the result shows that RESOLVE is the only surrogate model that successfully approximates gravitational-wave paleontology of binary black holes with proper statistical coverage. By evolving the binary black hole formation efficiency into a likelihood function, we produced a credible intervals for key astrophysical parameters including metallicity, envelope efficiency, $\sigma_{BH}$, and $\sigma_{NS}$. These credible intervals can be used by astronomers to validate existing theories or develop new ones regarding the formation, lives, and deaths of stars across cosmic time and the pivotal role they play in shaping our Universe.

Given that high dimensionality and prohibitive simulation costs represent the two principal challenges in Gravitational Wave Paleontology, RESOLVE establishes a framework to address both problems simultaneously. Our future work will proceed along two complementary directions: first, we aim to enhance RESOLVE by systematically addressing the limitations discussed in Section~\ref{section:limitation}; second, we plan to apply the refined RESOLVE framework to understand other properties of binary systems. This work demonstrated how AI can yield scientific conclusions that can be directly used by the astronomical community, fostering a productive interdisciplinary synergy that advances both fields.

\bibliography{neurips}
\bibliographystyle{plainnat}

\appendix
\section{The Rare Event Problem}\label{sec:theory}
This section aims to adapt the Rare Event Design problem in~\citep{schuetz2025resum} to binary black hole collision simulation. Supposed we simulated $N$ independent binary systems in COMPAS. For each system $i$, The outcome of $X_i$ is either 1, indicating that the $i^{th}$ pair of stars formed black hole pairs that collided and emitted a detectable gravitational wave, or 0, indicating that the binary system did not form colliding black holes. 
If $m$ collision happened in $N$ simulated binary star systems, the formation efficiency $\epsilon$ can be defined as:
\begin{equation}
    \epsilon = \frac{m}{N} = \frac{\sum_{i=1}^{N} X_i}{N}
\end{equation} 
Let $\boldsymbol{\theta}$ denote the physics parameter of interest and $\boldsymbol{\phi}$ the nuisance parameters. The value of $\epsilon$ may depend on both $\boldsymbol{\theta}$ and $\boldsymbol{\phi}$. Since our primary interest is on $\boldsymbol{\theta}$, we simulate $N$ events where all events share the same value of $\boldsymbol{\theta}$ but have different, randomly-sampled values of $\boldsymbol{\phi}$.

The number of collisions \( m \) follows a binomial distribution with the probability \( e(\boldsymbol{\theta}, \boldsymbol{\phi_i}) \). The high computational cost arises from the Rare Event Condition: since \( e(\boldsymbol{\theta}, \boldsymbol{\phi_i}) \) is intrinsically very small, the number of triggered signal $m$ becomes negligible compared to $N$ ($m\ll n$). This means \( m \) can be approximated by a Poisson distribution as $m \sim \text{Poisson}\left(N\bar{t}(\boldsymbol{\theta})\right)$.  Where $\bar{t}(\boldsymbol{\theta})$ is the expected triggering probability marginalized over the nuisance parameter $\boldsymbol{\phi}$:

\begin{equation}
    \bar{e}(\boldsymbol{\theta}) = \int e(\boldsymbol{\theta},\boldsymbol{\phi}) g(\boldsymbol{\phi}) d\boldsymbol{\phi}=\prod_i\int e(\boldsymbol{\theta}, \boldsymbol{\phi}_i) g(\boldsymbol{\phi}_i) d\boldsymbol{\phi}_i
    \label{eq:expected_trig_prob}
\end{equation}
$g(\phi)$ is a predefined uniform distribution where we sampled $\phi_i$ from. However, the analytical form of $e(\theta,\phi)$ is unavailable, and direct evaluation of $e(\boldsymbol{\theta},\boldsymbol{\phi})$ is impossible neither. We only have access to $X_i$, generated via:
\begin{equation}
    X_i = \text{Bernoulli}[t(\boldsymbol{\theta},\boldsymbol{\phi})]
\end{equation}

When $N$ becomes large, according to the central limit theorem, the formation efficiency $\epsilon$ will follow a normal distribution with symmetric, well-defined statistical uncertainties $\bar{e}(\theta)/N$; As $N\xrightarrow[]{}+\infty$, $\epsilon$ will asymptotically approximate $\bar{e}(\theta)$ with statistical uncertainties approaching 0. But When \( N \) is small, \( m \) has a higher variance; \( \epsilon = \frac{m}{N} \) can no longer be approximated with a normal distribution. In other words, $\epsilon$ will only takes on a discrete set of values, \( \epsilon \in \left\{ \frac{0}{N}, \frac{1}{N}, \dots, \right\} \).

\section{COMPAS Binary Population Synthesis}
\label{sec:COMPAS}
We use the open-source \ac{COMPAS}\footnote{\url{compas.science}. Code available at \url{https://github.com/TeamCOMPAS/COMPAS}.} binary population synthesis code to simulate the lives of massive binary stars from birth to death.
The primary scientific objective of \ac{COMPAS} is to understand the \acp{BH} and \acp{NS} whose mergers emit the gravitational waves that are detected today.
By simulating the lives of large populations of massive stars, the software is used to forward model observed compact object mergers, helping to constrain the population of compact objects, their massive stellar progenitors, and the physics that underlies stellar and binary evolution.

\ac{COMPAS} models binary events using simple parametrized prescriptions for isolated stellar and binary evolution, allowing it to compute the complete evolution of a binary in $\sim 10$ milliseconds on a typical laptop.
Other popular codes that use similar approaches to population synthesis include (but are not limited to) MOBSE \citep{Giacobbo_2018a, Giacobbo_2018b}, COSMIC \citep{Breivik_2020}, StarTrack \citep{Belczynski_2002, Belczynski_2008, Belczynski_2020}, and POSYDON \citep{Fragos_2023, Andrews_2024}.
\citet{compas_team_compas} gives an in-depth description of \ac{COMPAS}'s methodology and implementation, which we describe below.

Single stars constantly undergo nuclear fusion as they evolve, driving changes to their properties and structure, causing phenomena like stellar winds, and triggering events like supernovae.
To handle stellar evolution, \ac{COMPAS} uses the evolutionary formulae for properties including mass, metallicity, radius, and luminosity from \citet{Hurley_2000} which are based on stellar models from \citet{Pols_1998}.
The stellar tracks provided by \citet{Hurley_2000} segment the evolution of massive stars according to their evolutionary phases like the main sequence, Hertzsprung gap, giant branch, and more.
Once stars run out of thermonuclear fuel, \ac{COMPAS} uses a variety of class-specific prescriptions to model how they undergo a supernova and leave behind a stellar remnant either consisting of a white dwarf, \ac{NS}, or \ac{BH} based on their pre-supernova mass.
Stellar remnant evolution is then handled using separate sets of formulae for each remnant type.

As binaries evolve, they often interact through means beyond gravitational attraction, leading to changes in the component properties and orbit.
\ac{COMPAS} parametrizes binary events using their component masses, separation, and eccentricity, and uses prescriptions to account for mass loss, stable mass transfer, unstable mass transfer, supernovae, and stellar contact or mergers.
The impacts of gravitational radiation are only considered by \ac{COMPAS} after both binary components are compact objects, at which point binary eccentricity and separation decrease according to the point-mass approximation from \citet{Peters_1964}.
If gravitational radiation emission causes a binary to inspiral before Hubble Time (the current age of the Universe), it is considered a merger event.

To evolve populations of binaries, users provide \ac{COMPAS} with distributions for initial binary properties including component masses, separation, eccentricity, and metallicity.
The distributions of initial binary properties are meant to reflect what is observed in the Universe so that users can synthesize populations that represent nature and give insight into sampled subsets.
\ac{COMPAS} also provides several postprocessing scripts for calculating important astrophysical quantities using simulation results.
Some of the metrics that \ac{COMPAS} scripts can calculate include the efficiency of compact object formation and the rates of compact object coalescence and gravitational wave detection.

\section{COMPAS Training Set Parameter Choices \label{app:param_choices}}
In this section, the specific parameters for $\boldsymbol{\theta}$ and $\boldsymbol{\phi}$ chosen from the parameter space of COMPAS is outlined. The tables below gives an overview of the physical parameters of interest for $\boldsymbol{\theta}$. The list of parameters chosen for $\boldsymbol{\phi}$ are:
\begin{center}
\begin{tabular}{||c|c|c||}
\hline
CH on MS(1) & CH on MS(2) & Eccentricity@ZAMS \\
Equilibrated At Birth & Evolution Status & Mass@ZAMS(1)\\ Mass@ZAMS(2) & Merger & Merger at Birth \\
Metallicity@ZAMS(1) & Metallicity@ZAMS(2) & Omega@ZAMS(1)\\ Omega@ZAMS(2) & PO CE Alpha & PO LBV Factor \\
PO Sigma Kick CCSN BH & PO Sigma Kick CCSN NS & PO Sigma Kick\\
ECSN & PO Sigma Kick USSN & PO WR Factor \\
SN Kick Magnitude Random N & SemiMajorAxis@ZAMS & Stellar Type(1)\\
Stellar Type(2) & Stellar Type@ZAMS(1) & Stellar Type@ZAMS(2)\\ \hline
\end{tabular}
\end{center}

\section{The LEGEND Detector Design Dataset}\label{app:LEGEND_data}
The LEGEND detector design dataset is curated to find the optimal design of a neutron moderator that slows down and blocks external neutron background from entering the sensitive region. Given a specific detector design, the dataset runs GEANT4 simulations with $N$ neutrons with random initial energy, position and momentum, and count the number of neutrons that enter the sensitive region as $m$. The design metric is calculated as $y_{Raw}=m/N$. The parameter of interest $\boldsymbol{\theta}$ include 5 design parameters that control the shape of the neutron moderator, while the nuisance parameters $\boldsymbol{\phi}$ are the initial energy, position and momentum of each simulated neutrons.

This dataset contains both HF and LF simulations. The primary difference between HF and LF are (1) the number of neutrons simulated and (2) the physics mechanism behind neutron production. Each HF simulation trial requires 170 CPU hours, while each LF simulation trial only requires 15 CPU minutes. The training datasets contains 310 LF simulation trials and 10 HF simulation trials, while the validation dataset contains 100 out-of-sample HF simulation trials. 

After training, the surrogate model will predict $\hat{y}\pm\hat{\sigma}$ from the given $\boldsymbol{\theta}$ of each out-of-sample HF simulations. The means square error is calculated by averaging $(\hat{y}-y_{Raw})^2$ over the 100 trials, while the 1/2/3$\sigma$ coverage is calculated by counting the number of trials where $y_{Raw}$ fall within $\hat{y}\pm1/2/3\times\hat{\sigma}$, respectively. As shown in Table~\ref{tab:benchmark}, the MFGP algorithm (Trial 1) is significantly undercovered. The RESuM model (Trial 2) and the RESOLVE model (Trial 3) both achieved proper statistical coverage on the LEGEND detector design dataset, while RESOLVE outperforms RESuM on the MSE metric. This means that that RESOLVE achieves better prediction accuracy while maintaining proper statistical coverage.

\section{Converting Formation Efficiency to Gravitational Wave Rate}\label{app:epsilon_to_y}
Note that the emulated formation efficiency $\hat{\epsilon}$ represents the number of binary black hole collisions given $N$ simulated binary star systems. This unitless efficiency is different from the expected event rate $\hat{y}$, which has the the unit of number of expected collisions per unit of co-moving volume per year. To obtain $\hat{y}$, we need to perform another step of conversion listed in the following equation:
\begin{equation}
    \hat{y} = s\epsilon_r \hat{\epsilon}
    \label{eqn:rate_conversion}
\end{equation}
Where $s$ is the local star formation rate density, which is $1\times10^7$ times the mass of the sun per co-moving cubic Gigaparsec per year~\citep{Madau:2014bja}; $\epsilon_r$ is a reference efficiency of the mean mass evolved in our simulations which equals $1.57133427\times10^{-6}/0.005$ events per solar mass based on \citep{compas_team_compas}. Since both numbers are constant, Equation~\ref{eqn:rate_conversion} simplifies to a constant number multiplication: $\hat{y} = 314.266854 \cdot \hat{\epsilon}$.

\section{A Brief Summary of Gravitational Radiation from Rotating events \label{app:gravity_radiation_rotation}}
Following the derivation from Weinberg \cite{steven_weinberg_gravitation_1972}, we consider a stellar body with a mass density denoted by $\rho(\boldsymbol{x'})$. Assuming this body is rigidly rotating about a fixed axis with a constant angular velocity, the power radiated from this event is given by equation (10.5.22) \cite{steven_weinberg_gravitation_1972}:
\begin{equation}
    P(2\Omega)\propto{\Omega^6I^2e^2}
\end{equation}
Where $\Omega$ is the rotation frequency. Let $I_{ij}$ ($i,j\in \{1,2,3\}$) be the moment of inertia tensor for this event. Then $I$ shown above is given by $I=I_{11}+I_{22}$ and:
\begin{equation}
    e=\frac{I_{11}-I_{22}}{I}
\end{equation}
Now, if the rigid body is simply rotating at the center of the rotation axis, then $e=0$, and thus the radiated power is $0$. However, for a event where a body is rotating about the center axis (thus the circular symmetry is broken), both $I$ and $e$ are non-zero, and there is therefore non-zero radiated power.
\newparagraph
A event such as a binary black hole event may admit gravitational waves whereas a single-star event may not, even if it is rotating. Non-zero gravitational wave contributions may arise from particle collisions inside the stellar body, but these effects may pale in comparison to the radiation caused by a binary event.

\section{Bayesian Markov Chain Monte Carlo (MCMC) \label{sec:bmcmc}}
In particle physics and other experimental contexts, Bayesian inference is oftentimes employed to compare data to theoretical models. In this section, we give a brief review of Bayesian Markov Chain Monte Carlo (MCMC) and outline an example of its application to particle physics.
\newparagraph
\emph{Bayesian} statistics, as compared with \emph{frequentist} statistics, uses data from an experiment and a model to give a likelihood as to whether the data matches the model. The method revolves around \emph{Bayes' Rule}. When there are $n$ discrete possible outcomes, this is \cite{john_pitman_probability_1993}:
\begin{equation}
    P(B_i|A)=\frac{P(A|B_i)P(B_i)}{\sum_i^n P(A|B_i)P(B_i)}
\end{equation}
This principle, which describes how conditional probabilities may be "flipped" can be easily extended to a statement on theory and experiment. Let $M$ be a theoretical model with a set of model parameters $\theta_M$. This model is to characterize a collection of data, denoted $D$. The probability that the  model parameters are correct given some data and a model is given by \cite{speagle2020conceptualintroductionmarkovchain}:
\begin{equation}
P(\boldsymbol{\theta)\coloneq}P(\boldsymbol{\theta}_M|\boldsymbol{D},M)=\frac{P(\boldsymbol{D}|\theta_M,M)P(\boldsymbol{\theta}_M|M)}{\int_{\theta_M}P(\boldsymbol{D}|\theta_M)P(\boldsymbol{\theta}_M|M)d\theta_M}
\end{equation}
For completion, $P(\boldsymbol{D}|\theta_M,M)$ is the probability that the data came from a model with a chosen set of parameters, and $P(\boldsymbol{\theta}_M|M)$ is the probability that the model parameters are true given a model.
\newparagraph
In the Bayesian language, one may want to maximize the probability that the model parameters (and the model) accurately describe the experimental data. However, since the "true" model parameters may not be known, a surrogate scenario may be used, where the \emph{expected loss} may be minimized. That is, if the true value of $\boldsymbol{\theta}$ is known, different values of parameters may be suggested, $\boldsymbol{\hat{\theta}}$, and a \emph{loss function}, $L_P(\hat{\theta},\theta)$, can penalize this suggestion by comparing with the true value. However, since the true parameter might not be known, one can marginalize over all possible parameters by:
\begin{equation}    
L_P(\hat{\theta})\coloneq \mathbb{E}_P[L_P(\hat{\theta}|\theta)]
\end{equation}
And find the best value of $\boldsymbol{\theta}$ by minimizing the expected loss:
\begin{equation}
    \operatorname*{arg\,min}_{\hat{\theta}\in\theta}~\mathbb{E}_P[L_P(\hat{\theta}|\theta)]
    \label{eqn:expected_loss}
\end{equation}
While this may be formally straightforward, it is oftentimes computationally expensive to navigate over all possible model parameters, and this problem can scale as an NP\footnote{Non-deterministic polynomial time}-complex problem.
\newparagraph
However, there are ways to improve the complexity of the problem. One way is to traverse the joint parameter space at (potentially) uneven intervals. That is, the expected loss may be approximated by a discrete sum that resembles a Riemannian sum:
\begin{equation}
    \mathbb{E}_P[L_P(\hat{\theta}|\theta)]=\int L(\boldsymbol{\theta})P(\boldsymbol{\theta})d\theta\rightarrow \sum_{i=1}^n L(\boldsymbol{\theta}_i)P(\boldsymbol{\theta}_i)\Delta \boldsymbol{\theta}_{i,i+1}
\end{equation}
Where the term $\Delta \boldsymbol{\theta}_{i,i+1}$ denotes the spacing between the set of $n$ model parameters. It can be shown \cite{wiegand_kish_1968} that the most optimal way to space the intervals is such that the corresponding spacing on the posterior distribution, $P(\boldsymbol{\theta})$, is minimal where the posterior is greatest in values and maximal where it is lowest. In this way, the model parameters that marginally contribute the least to the posterior distribution are "skipped over" the fastest, and the parameters that contribute the most are tediously searched over. Stated pictorially, this is:
\begin{equation}
    \text{parameter spacing} \propto \frac{1}{\text{posterior distribution}}\rightarrow Q(\boldsymbol{\theta})\propto \frac{1}{\Delta \boldsymbol{\theta}(\boldsymbol{\theta})}
\end{equation}
This strongly motivates (and exactly produces) a new distribution, called the \emph{proposal distribution}, $Q(\boldsymbol{\theta})$. In the background, this has changed the parameter search problem introduced in Equation \ref{eqn:expected_loss} to be:
\begin{equation}
    \mathbb{E}_P[{}\cdot{}]\rightarrow \mathbb{E}_Q[{}\cdot{}]
\end{equation}
Using this formalism, the problem statement shifts to a problem regarding the distribution of parameter space rather than the posterior distribution itself. \emph{Monte Carlo methods}, which are computational techniques involving sampling from a discrete set of values as a stand-in for the distribution in question, can now be employed on the distribution of possible intervals.
\newparagraph
Having motivated the use of Monte Carlo methods in searching parameter spaces in Bayesian inference problems, we discuss the Markov Chain Monte Carlo technique. This approach seeks to create an optimal distribution over parameter space by generating samples from a Markov Process. A Markov Process, in this context, is a random walk in parameter space that produces a set of parameters explored. That is, the \emph{chain} of parameters explored is described by $n$ random steps (iterations):
\begin{equation}
    \{ \boldsymbol{\theta}_1,...,\boldsymbol{\theta}_n\}
\end{equation}
From this chain of parameter values, a probability density is mimicked by simply counting the number of parameter values from the chain inside a volume element and dividing by the total number of values generated: note that this volume element is the same used in computing the posterior distribution. This equation is given in Equation 44 in Reference \cite{speagle2020conceptualintroductionmarkovchain}:
\begin{equation}
    \int_{\theta\in \delta_{\theta}}  P(\boldsymbol{\theta})d\theta\approx n^{-1} \sum_{j=1}^n \boldsymbol{\mathbbm{1}}[\boldsymbol{\theta_j} \in \delta_{\theta} ]
    \label{eqn:mcmcdensity}
\end{equation}
In the above equation, $\boldsymbol{\mathbbm{1}}[{}\cdot{}]$ is the indicator function and $\delta_\theta$ is the volume element. Since this stand-in density is inversely proportional to the posterior distribution, simulation can now be made over parameter space instead of the posterior using a Markov Chain process. One popular implementation of a Markov Chain Monte Carlo technique is the Metropolis-Hastings Algorithm \cite{metropolis_equation_1953}. In this method, the chain of parameters is produced (at least conceptually) in the following way:
\begin{enumerate}
    \item In the chain of parameters, suggest a new parameter based on the current parameter in the list using a \emph{proposal distribution}, $Q(\boldsymbol{\theta_{i+1}}|\boldsymbol{\theta_i})$
    \item This new parameter is accepted with some transition probability or rejected with a potentially different probability
    \item Once the probability of accepting a new parameter is equal to rejecting the parameter\footnote{This idea stems from \emph{detailed balance}, which coincidentally has strong connections in equilibrium statistical mechanics}, the optimal proposal distribution has been found
\end{enumerate}
This proposal distribution emulates the behavior of the density function described by Equation \ref{eqn:mcmcdensity}, which means it has direct relation to the posterior distribution.

\section{Bayesian MCMC in Particle Physics}\label{app:Bmcmc_particle_physics}
In Appendix \ref{sec:bmcmc}, we described how Bayesian MCMC changes the language of model tuning from one involving the posterior distribution to a a search in parameter space. In this section, we describe how such an approach is commonly used in particle physics applications, and motivate its use in an astronomical setting. Using the principles of Bayesian MCMC, experimental particle physicists test theoretical models by optimizing the model's parameters. However, finding the optimal parameter space distribution is merely the beginning of the analysis. After such distribution is found, what is oftentimes most useful to physicists is the marginal distribution over different parameters, as understanding high-dimensional spaces may not be as intuitive. Consider the following figure from \cite{speagle2020conceptualintroductionmarkovchain}:
\begin{center}
    \includegraphics[scale=0.35]{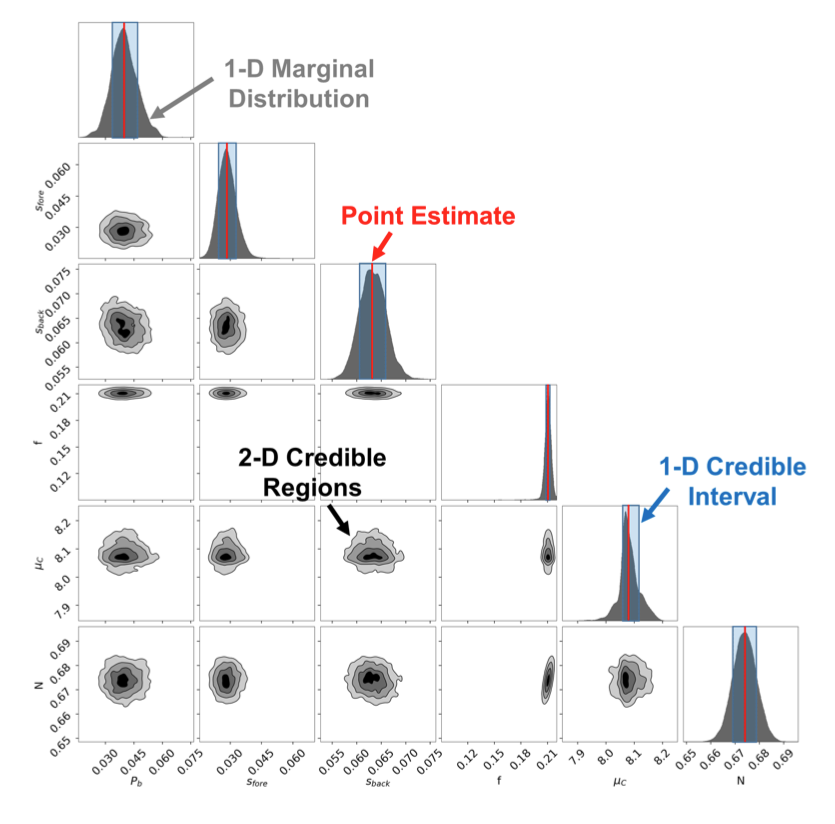}
    \label{fig:speagle_mcmc}
\end{center}
This generic graph shows the marginal distributions over several parameters, namely $\theta=\{P_b,s_{fore},s_{back},f,\mu_c,N\}$. Viewing this set of graphs as a matrix, the diagonal elements (shown with red lines) are the 1-dimensional distributions of the parameters: for example, the top left graph is a 1-dimensional marginal distribution of $P_s$ (along the x-axis). The "off-diagonal" graphs are the joint marginal distributions between \emph{two} parameters. This explains why the graphs are shown with contours, as a probability density between two variables is naturally three-dimensional. Filling in all the "off-diagonal" graphs, two-variables at a time may be compared, which has a much easier interpretation for how different model parameters vary together. Having motivated why marginalizing the MCMC parameter distribution gives insight into how model parameters vary together, we now turn to an example of their application in particle physics.
\newparagraph
CERN, which is the world's leading high energy particle physics experiment, is home to several important scientific advancements, including the discovery of the Higgs Boson \cite{higgs_boson}. To explore high energy physics, particles are accelerated to ultra-relativistic\footnote{Speeds close to the speed of light} speeds and collide into other particles to expose physics processes at the subatomic scale. This process, broadly described by \emph{scattering theory}, can quickly demand a large amount of model parameters that must be tested experimentally. Many of CERN's experiments must, therefore, explore large parameter spaces to verify the experimental data. MCMC is a natural choice for analyzing experimental data. One example of CERN using MCMC in their analysis may be found from \cite{coccaro_dnnlikelihood_2020}. In this work, a new framework for likelihood estimation in large parameter spaces is presented. Part of the model features sampling from an MCMC process, and in Figure 9 a plot similar to Figure \ref{fig:speagle_mcmc} is shown \cite{coccaro_dnnlikelihood_2020}:
\begin{center}
    \includegraphics[scale=0.25]{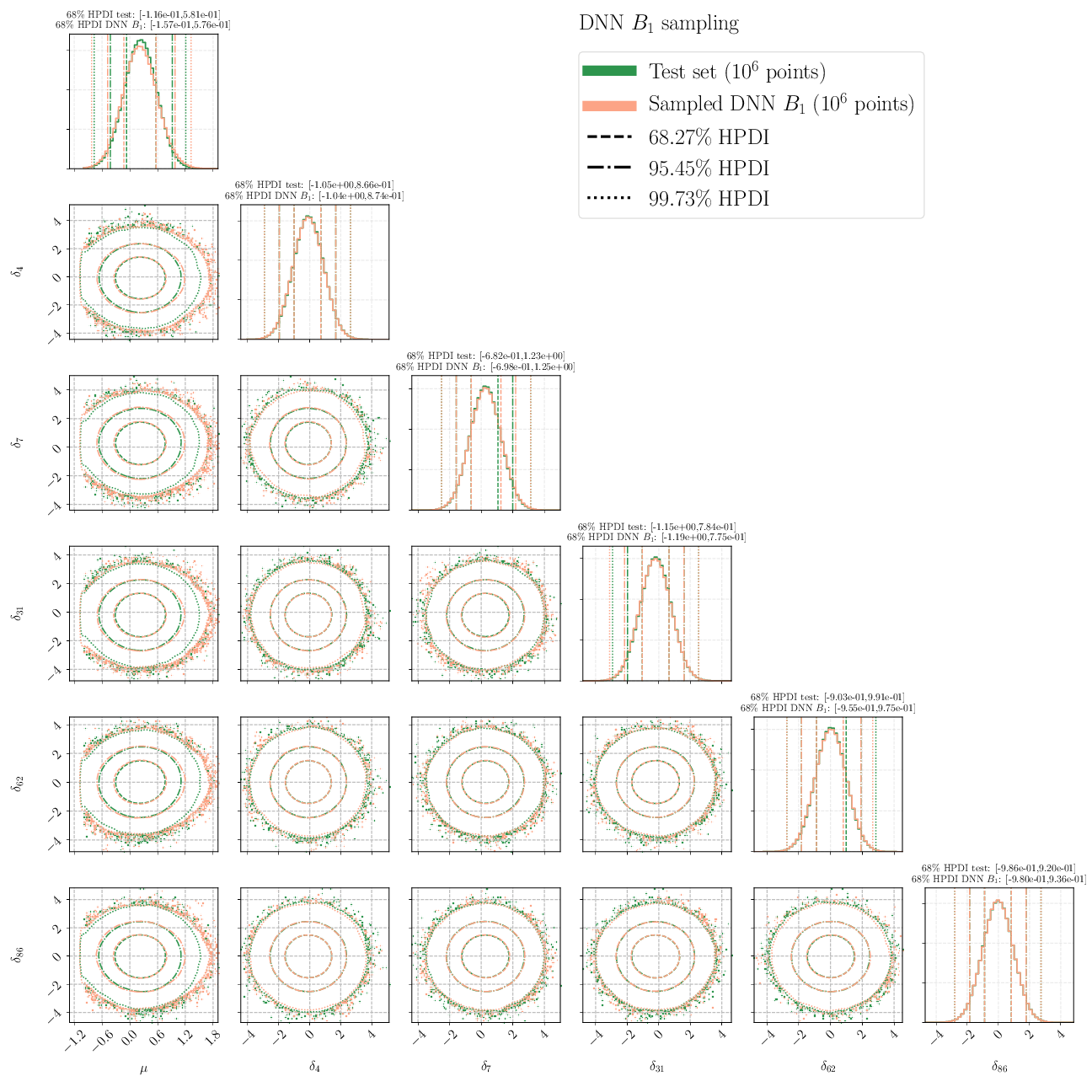}
    \label{fig:cern_mcmc}
\end{center}
In this figure, marginal distributions are shown for various model parameters in a machine learning model. While this paper is certainly not the only use of MCMC in CERN experiments, it does underscore the basic principles of MCMC in high energy physics. Moreover, CERN is not the only experiment in high energy physics that employs MCMC.

\section{Model Implementation and Diagnostics}\label{sec:diagnostics}

Accurate surrogate modeling in a multi-fidelity setting requires balancing model complexity with the availability of data across fidelity levels. To achieve this, we adopt a targeted polynomial order selection strategy, employ Bayesian inference with carefully chosen priors, and assess model quality through a comprehensive set of diagnostics. All core modeling components are implemented in Python, using the \texttt{chaospy}, \texttt{scikit-learn}, \texttt{PyMC}, and \texttt{aesara} libraries.

\subsection{Polynomial Order Selection}

The polynomial order for the low-fidelity model is selected using \( K \)-fold cross-validation. For a predefined set of candidate degrees \( d \), we generate the corresponding multivariate orthonormal polynomial basis using the \texttt{chaospy} package and split the low-fidelity training data into \( K \) folds. The surrogate model is trained on \( K - 1 \) folds and evaluated on the remaining fold by computing the mean squared error (MSE). This procedure is repeated across all folds, and the average MSE is used to estimate the generalization error for each degree. The optimal degree \( d^* \) is then selected as the one minimizing the average cross-validated MSE.

To mitigate overfitting and enforce sparsity in the regression coefficients, we apply LASSO regression with automatic cross-validation using \texttt{scikit-learn}'s \texttt{LassoCV}. This step helps identify the most relevant polynomial terms for the surrogate expansion.

Due to the sparsity of higher-fidelity data, we fix the polynomial order for the discrepancy model (used to correct the lower-fidelity prediction) to 1. This linear model captures leading-order deviations without overfitting and ensures stable estimation when only a limited number of high-fidelity samples is available.

\subsection{Bayesian Inference}

Bayesian inference is performed using Hamiltonian Monte Carlo (HMC), specifically the No-U-Turn Sampler (NUTS), implemented via the \texttt{PyMC} probabilistic programming framework. This allows for efficient exploration of the high-dimensional posterior space and provides full uncertainty quantification over model parameters. Tensor algebra operations needed in the model specification, such as linear transformations and inner products, are expressed symbolically using \texttt{aesara}, which underpins \texttt{PyMC}'s computational backend.

Priors are specified individually for each fidelity level, allowing the model to reflect varying levels of uncertainty and data availability across fidelities.

\subsection{Prior Distributions}

We place prior distributions on the unknown model parameters to encode our assumptions and promote regularization:

\begin{itemize}
    \item \textbf{PCE coefficients:} Each coefficient \( c_{j/k}^{(f_i)} \) is assigned a zero-mean Gaussian prior \( \mathcal{N}(0, \tau_{j/k}^2) \), encouraging sparsity and penalizing irrelevant terms.
    
    \item \textbf{Noise variance:} The noise standard deviation \( \sigma^{(f_i)} \) is modeled with a Half-Normal prior, \( \sigma^{(f_i)} \sim \text{HalfNormal}(\lambda) \), where \( \lambda \) is set to the empirical standard deviation of the high-fidelity data. This reflects the belief that noise is strictly positive and typically small.
    
    \item \textbf{Fidelity scaling coefficient:} For each fidelity level \( f_i > \text{LF} \), we introduce a scaling coefficient \( \rho^{(f_i)} \sim \mathcal{N}(\mu, \sigma_{\rho^{(f_i)}}^2) \). Rather than fixing the mean \( \mu = 1 \), we determine it by minimizing the MSE between the lower-fidelity and the higher-fidelity training data. This optimization allows \( \rho^{(f_i)} \) to reflect empirical discrepancies between fidelity levels rather than enforcing a fixed prior belief.
\end{itemize}

These priors are flexible and can be adapted to encode stronger or weaker assumptions, depending on domain knowledge and the observed data characteristics.

\subsection{Model Comparison and Diagnostics}

To evaluate model quality and support model comparison, we compute the Widely Applicable Information Criterion (WAIC) and Leave-One-Out Cross-Validation (LOO-CV) using Monte Carlo estimates from posterior samples obtained via \texttt{PyMC}. These criteria provide insight into out-of-sample predictive performance and are complemented by a robust set of additional diagnostics:

\begin{itemize}
    \item \textbf{Calibration:} Posterior predictive intervals at \( \pm 1\sigma \), \( \pm 2\sigma \), and \( \pm 3\sigma \) are visually inspected to assess coverage and uncertainty calibration.
    
    \item \textbf{Predictive performance:} We report the mean squared error (MSE), expected log predictive density (\( \text{elpd}_{\text{loo}} \)), its differences across competing models (\( \text{elpd}_{\text{diff}} \)), and associated standard errors. The effective number of parameters is estimated via \( p_{\text{loo}} \) and \( p_{\text{waic}} \).
    
    \item \textbf{Empirical coverage:} We compute empirical coverage probabilities at the \( \pm 1\sigma \), \( \pm 2\sigma \), and \( \pm 3\sigma \) levels to evaluate the reliability of posterior intervals.
\end{itemize}

\subsection{Sampling Diagnostics}

We monitor the efficiency and convergence of the MCMC sampling process using several diagnostics, available through \texttt{PyMC} and its companion library \texttt{ArviZ}:

\begin{itemize}
    \item \textbf{Bayesian Fraction of Missing Information (BFMI):} Assesses whether the HMC algorithm adequately explores the energy landscape.
    
    \item \textbf{Potential scale reduction factor (\( \hat{r} \)):} Checks for convergence across multiple chains.
    
    \item \textbf{Effective sample size:} Inferred from posterior variance and stability of interval estimates.
\end{itemize}

We also examine the Pareto shape parameters and posterior variance of log predictive densities to identify influential observations that may compromise the reliability of WAIC and LOO-CV. These diagnostics help ensure robust inference and guard against overconfidence in model selection based on information criteria alone.

\section{Model Behavior with Respect to Envelope Parameter}\label{sec:alpha_behavior}

\begin{figure}[hb!]
    \centering
    \includegraphics[width=0.98\linewidth]{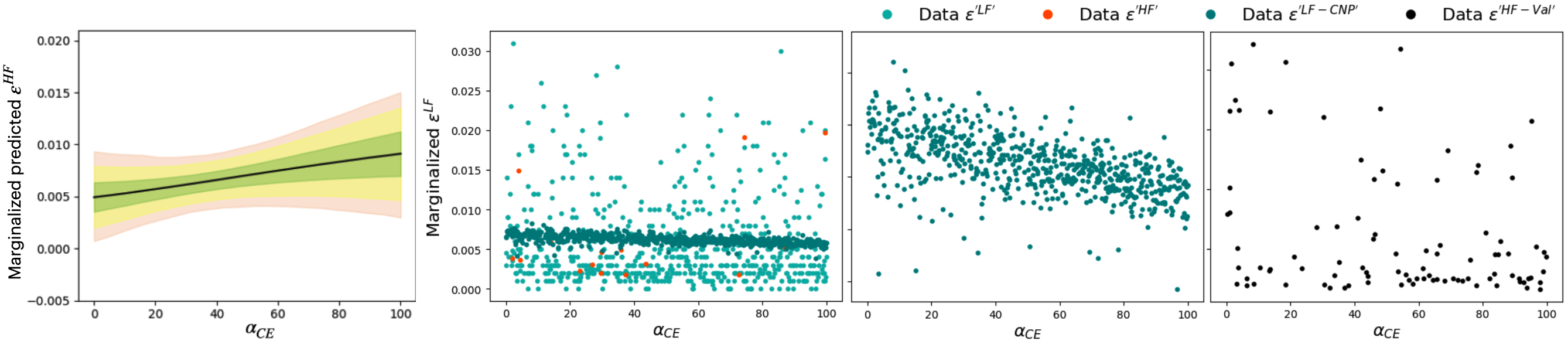}
    \caption{Marginalized model predictions and data for the merger efficiency \( \epsilon \) as a function of the common-envelope parameter \( \alpha_{\mathrm{CE}} \). \textbf{Left:} Predicted HF response \( \epsilon^{HF} \) with 1\(\sigma\), 2\(\sigma\), and 3\(\sigma\) uncertainty bands. \textbf{Middle-left:} LF and HF training data used to fit the surrogate model. \textbf{Middle-right:} CNP-based LF predictions used in the multi-fidelity model. \textbf{Right:} HF validation data.}
    \label{fig:envelope}
\end{figure}

In Section~\ref{section:experiment}, we reference an observed discrepancy between the surrogate model’s predictions and the smoothed low-fidelity trend for the parameter $\alpha_{\text{CE}}$. Figure~\ref{fig:envelope} compares the marginalized low-fidelity outputs ($\epsilon^{LF}$, light teal), their CNP-smoothed counterparts ($\epsilon^{LF\text{-CNP}}$, dark teal), and both the high-fidelity training data (red) and high-fidelity validation data (black). While the raw low-fidelity outputs appear noisy and structureless due to their scale and variance, the CNP reveals underlying trends by effectively denoising the signal. The surrogate model, shown in Figure~\ref{fig:envelope}, incorporates these smoothed estimates but ultimately prioritizes the high-fidelity training points—15 in total—which it assumes best approximate the ground truth. This weighting explains the model’s upward trend with increasing $\alpha_{\text{CE}}$, despite the CNP predicts a very slight decrease in the formation rate. However, when considering the high-fidelity validation points—unseen during training—the upward trend appears much less pronounced and more scattered. This suggests that the training HF dataset may introduce a slight bias due to its sparsity, leading the model to infer structure that does not generalize. Consequently, in data-scarce regimes, limited and potentially unrepresentative high-fidelity samples can strongly influence the surrogate model, underscoring the importance of careful HF selection and the need for uncertainty-aware modeling strategies.

To mitigate the model’s sensitivity to such biases, particularly for $\alpha_{\text{CE}}$, several approaches can be considered. One option is to weaken the prior on the residuals between low- and high-fidelity levels, thereby reducing the model’s tendency to overfit to a small HF set, though at the potential cost of increased predictive error and reduced coverage. Alternatively, structured priors can be placed directly in the PCE by shrinking the coefficients of basis terms involving $\alpha_{\text{CE}}$, encoding prior knowledge or skepticism about its influence. Hierarchical priors offer an even more flexible solution by allowing the model to infer the relevance of $\alpha_{\text{CE}}$ directly from the data. As visible in the right region of Figure~\ref{fig:envelope}, the model’s uncertainty increases where high-fidelity data are lacking, suggesting that targeted sampling in such regions could both reduce predictive uncertainty and correct potential biases in the inferred trend. A similar active learning approach was used in the RESuM method \cite{schuetz2025resum}, which also employed integrated variance reduction to guide high-fidelity simulations in a data-efficient manner. In this work, we directly benchmark against RESuM and compare performance under similar data-scarce conditions.

\end{document}